\documentclass[11pt]{article}
\usepackage{graphicx}
\usepackage{subfigure,amssymb,amsmath}
\usepackage{cite,color,url}
\usepackage{bm}
\usepackage{float}
\usepackage{jheppub}
\def\Vev{{\it vev}}
\def\Vevs{{\it vevs}}
\def\321{$\rm SU(3)_C\times SU(2)_L\times U(1)_Y$}

\def\10{SO(10)}
\def\sg{\tilde{g}}
\def\sq{\tilde{q}}

\def\sl{\tilde{l}}

\def\lspone{\widetilde\chi_1^0}
\def\mlspone{m_{\lspone}}

\def\sigmasip{\sigma_{\chi p}^{SI}}

\def\sigmachionesip{\sigma_{\lspone p}^{SI}}

\def\lsim{\ ^<\llap{$_\sim$}\ }

\def\issue(#1,#2,#3){{\bf #1}, #2 (#3)}%AIP format!Vol,page(Year)
\def\PREP(#1,#2,#3){Phys.\ Rep. \issue(#1,#2,#3)}
\newcommand{\stopr}{\tilde{t}_R}
\newcommand{\stopl}{\tilde{t}_L}

\newcommand{\sbottoml}{\tilde{b}_L}
\newcommand{\sbottomr}{\tilde{b}_R}

\newcommand{\hu}{H_u}
\newcommand{\hd}{H_d}
\newcommand{\msq}[1]{m^2_{#1}}
\newcommand{\lb}{\left(}
\newcommand{\rb}{\right)}
\newcommand{\yuk}[1]{y_{#1}}

\newcommand{\ms}[2]{m_{\tilde{#1}_{#2}}}

\newcommand{\MW}{M_{W}}

\newcommand{\gev}{\rm GeV}
\newcommand{\tev}{\rm TeV}
\def\lspone{\widetilde\chi_1^0}
\def\mlspone{m_{\lspone}}
\newcommand{\staur}{\tilde{\tau}_R}
\newcommand{\staul}{\tilde{\tau}_L}

\title{Exploring MSSM for charge and color breaking and other constraints in the 
context of Higgs@125 GeV.}

\author[]{Utpal Chattopadhyay,}
\author[]{Abhishek Dey}

\affiliation[]{Department of Theoretical Physics, Indian Association 
for the Cultivation of Science, 
2A \& B Raja S.C. Mullick Road, Jadavpur, 
Kolkata 700 032, India}

\emailAdd{tpuc@iacs.res.in}
\emailAdd{tpad4@iacs.res.in}

\abstract{
Exploring MSSM parameter space after the discovery of Higgs Boson at 125 GeV 
naturally demands large top-squark mixing or large trilinear coupling parameter 
$A_t$ in particular, so as to avoid excessively heavy squark, specially 
for the universal models like CMSSM. 
We study stability of electroweak symmetry breaking vacua in possible 
presence of deeper charge-color symmetry breaking minima within MSSM.    
Besides stable vacua, we consider scenarios characterized by the 
presence of global CCB minima, 
with SM like charge and color conserving vacuum, having stability over 
cosmologically large lifetime {(\it long-lived states)}. 
We allow vacuum expectation values for both stop as well as sbottom fields, since 
these belong to the third generation of sfermions with larger Yukawa couplings that have 
immediate effect on the tunneling time.  
Moreover, for large $\mu$ regions, radiative corrections to
Higgs boson mass from bottom-squark loop is quite significant.  
Regions of MSSM parameters space become viable for large $A_t$ and large $\mu$ 
 zones which are generically excluded via the traditional analytical CCB 
constraints. 
For a large value of $\tan\beta$,   
safe vacua associated with large values of $|\mu|$ and $|A_t|$ are 
predominantly long-lived and may be associated with relatively light stop masses. 
We also identify low $\mu$ regions associated with long-lived states.  
Both the above zones can be friendly to muon $g-2$ constraint. We also impose constraints from      
${\rm Br}(B \rightarrow X_s \gamma)$ and 
${\rm Br}(B_s \rightarrow \mu^+ \mu^-)$. We do the analysis for a 
moderate and a large $\tan\beta$.  We choose 
an example parameter point in the gaugino mass plane of $M_1$, $M_2$ that satisfies  
the dark matter constraints, basically a decoupled sector with respect to CCB.}

\begin{document}
\maketitle
\section{Introduction}
In the Standard Model (SM)\cite{SMrefs}
 of Particle Physics the electrically neutral component of 
the electroweak scalar doublet (Higgs) takes a
non-zero vacuum expectation value (\Vev) in the ground state 
leading to spontaneous symmetry breaking (SSB). This 
leads to generation of mass of $\rm SU(2)_L$ gauge
bosons and mass for fermions through Yukawa terms.  
Lorentz invariance of the vacuum prevents any object 
other than a Lorentz scalar from acquiring a non-zero \Vev.  
The only scalar present in SM is the Higgs scalar 
which is singlet under SU(3) color ($\rm SU(3)_C$).
The presence of physically equivalent continuum of degenerate 
minima in SM Higgs potential, enables one to define 
the unbroken U(1) generator as the electric charge. 
This along with unbroken $\rm SU(3)_C$ leads to charge and color conservation 
for the ground state of SM,~where the Higgs field acquires a non-vanishing \Vev.  
Supersymmetry (SUSY) that can potentially ameliorate the hierarchy 
problem associated with SM  is one of the most 
viable candidates for Beyond the Standard Model (BSM) 
physics~\cite{SUSYreviews1,SUSYreviews2,SUSYbook1,SUSYbook2}.  
In the simplest SUSY
extension of SM, namely the Minimal Supersymmetric Standard 
Model (MSSM)\cite{SUSYreviews1,SUSYreviews2,SUSYbook1,SUSYbook2,djouadi}, SM fermions and bosons  
are supplemented by bosonic and fermionic partners 
transforming under the same SM gauge group \321.  Thus there are 
new scalars like squarks ($\sq$) and sleptons ($\sl$) 
that are charged under $\rm {SU(3)_C}$ and $\rm U(1)_{EM}$.  
The full MSSM scalar potential may indeed have 
several minima where squarks or sleptons may additionally acquire non-zero \Vevs.  
Since the violation of charge and/or color quantum number
is yet to be observed, it is understood that the Universe at present is at a ground state which is
Standard Model like (SML), with only Higgs scalars acquiring \Vevs.
{\it A priori} it indicates that those parts of the multi-dimensional
parameter space corresponding to MSSM scalar potential 
that allow a deeper charge and color breaking (CCB) 
minima\cite{Casas:1995pd,LeMouel:2001ym,
AlvarezGaume:1983gj,Gunion:1987qv,Strumia:1996pr,Baer:1996jn,Brhlik:2001ni,Bordner:1995fh,Cerdeno:2003yt} should be 
excluded.  This puts severe constraints on the 
parameter space.  However, truly there is no reason to assume
that the present SML minima where the Universe rests 
is a true vacuum.  The Universe, in principle can 
rest in a local minima/false vacuum, provided the 
lifetime of this SML minima with respect to the decay time 
into a deeper CCB minima transition is 
cosmologically large (larger than the age of the Universe).  
The Universe is then said to reside in a {\it long-lived} state\cite{Kusenko_paper1,kusenko2,Brandenberger:1984cz,Riotto:1995am,Falk:1996zt}.  
Analytically minimizing MSSM potential containing large number of 
scalar fields is very difficult unless one considers 
simplifying assumptions.  This may even put more stringent 
constraints than what are actually required \cite{Kusenko_paper1,kusenko2,Cohen:2013kna}. 
Hence, for a given point in the multi-dimensional parameter space
it is important to check the existence
of any deeper CCB minima numerically as exhaustively performed in codes like 
{\tt Vevacious}\cite{Camargo-Molina:2013qva}
that in turn uses {\tt CosmoTransitions}\cite{Wainwright:2011kj}. In case 
such minima exist one should compute
the lifetime of the false vacuum and decide on the 
validity of the given point of parameter space 
depending on the computed lifetime. A parameter point which would 
either correspond to a stable or a long-lived vacuum state 
would be referred to have a safe vacuum.

Since the Higgs boson has been found to have a mass of around 
125~GeV \cite{HiggsDiscoveryJuly2012,Chatrchyan:2013lba} which 
is not very far from the upper limit of MSSM predicted value ($\sim 135$~\gev), it has become 
important to explore the phenomenological MSSM (pMSSM)\cite{Djouadi:1998di} 
parameter space that may give large 
radiative corrections to the Higgs boson mass and provides with a relatively lighter top squarks.  This may be handled 
by properly considering the trilinear coupling parameter $A_t$ and the 
Higgsino mixing parameter $\mu$, both of which may on the other hand be sensitive to 
the CCB constraints. After the Higgs boson is discovered, analyses 
have been performed considering the existence of long lived states both in 
Constrained MSSM (CMSSM) as in Ref.\cite{Camargo-Molina:2013sta} as 
well as in pMSSM context as worked in 
Ref.\cite{Chowdhury:2013dka,Blinov:2013uda,Camargo-Molina:2014pwa}. 
All the above works that probed CCB minima numerically, considered values of $\mu$ less than a TeV 
or so while exploring $A_t$ appropriately via satisfying the requirement of long-lived 
states and the Higgs mass constraint. In this work we probe the 
pMSSM parameter 
space in a wider area of $\mu-A_t$ plane for specific zones of $\tan\beta$, where $\tan\beta$ 
is the ratio of Higgs vacuum expectation values, and explore the possibility of 
long-lived states that would also satisfy phenomenological 
constraints from ${\rm Br}(B \rightarrow X_s \gamma)$,   
${\rm Br}(B_s \rightarrow \mu^+ \mu^-)$, muon $g-2$
and dark matter.  We will specifically explore the above 
scenario for large values of $\tan\beta$ that may have important effect 
on the existence of long-lived states.   We should mention 
here that both ${\rm Br}(B \rightarrow X_s \gamma)$ and    
${\rm Br}(B_s \rightarrow \mu^+ \mu^-)$ may have  
important characteristics for large 
$\mu$, $A_t$ and $\tan\beta$. While analyzing large $\mu$ scenarios 
we also allow non-vanishing \Vevs~ for third generation of 
scalar fields beyond 
top-squarks. We will additionally highlight the issue of radiative corrections to the Higgs boson mass 
from the bottom-squark and the tau-slepton sectors for large values of 
$\mu \tan\beta$ that could potentially reduce the mass of Higgs boson 
while $\mu$ is increased. 
 Furthermore, as we will see soon, validity of both large and small 
$\mu$ regions, with large $A_t$ may be highly interesting in relation to the muon $g-2$ result. 

This work which is done using {\tt Vevacious}\cite{Camargo-Molina:2013qva} 
is organized as follows.  In Section~2 we 
briefly discuss essential theoretical aspects of
CCB minima,~decay of false vacuum and its 
theoretical implication on MSSM.  In Section~3 
we present the results of our analysis as follows. 
First, we will discuss the results for low values of $\mu$ showing 
the conformity with past analyses and extend the work for a 
given large value of $\mu$. 
Thereafter, within the above section we will show the  
results of scanning over a wide region of  
$\mu-A_t$ plane for a moderate as well as for a large value of $\tan\beta$.
We also discuss the compatibility of our analysis with relevant low 
energy constraints like those from B-physics and cosmological constraints 
from neutralino dark matter\cite{dark_matter_ref}. 
We will further discuss the issue of 
muon $g-2$ in the context of long-lived vacuum scenario, presenting also 
a few benchmark points.  
Finally, we will conclude in Section-4.
\section{Aspects of CCB minima, Decay of False Vacuum and MSSM}
\label{ccbmssm}
The MSSM scalar sector consists of squarks and the sleptons 
and two Higgs doublets with opposite $\rm U(1)_Y$ hypercharge. 
The sfermions, charged under \321 may acquire non-zero \Vevs.  
This may result into the existence of potentially dangerous CCB minima 
that may lie below an SML vacuum.  
The rate of tunneling from SML false vacuum to such CCB true vacuum 
is roughly proportional to  
$e^{-a/y^2}$, 
where $a$ is 
a constant and $y$ is the Yukawa coupling, signifying a larger decay rate 
for enhanced Yukawa coupling\cite{Kusenko_paper1,kusenko2,LeMouel:2001ym}.  
Hence the third 
generation of sfermions will be the most important candidate in 
connection with the formation of potentially dangerous global minima. 
We do not consider the direction where a sneutrino 
$\widetilde{\nu}$
may acquire a \Vev,~ since the corresponding vacuum would conserve 
both charge and color.    
In this analysis we would limit ourselves on scenarios where only third 
generation of squarks  
may acquire \Vevs. 

\noindent
For simplicity, we now focus on the stop and Higgs fields of the  
MSSM scalar potential~\cite{SUSYreviews1,SUSYreviews2,SUSYbook1,SUSYbook2}.
\begin{align}
V = &\; \lb \msq{\hu} + \mu^2 \rb |\hu|^2 + \lb \msq{\hd} + \mu^2 \rb |\hd|^2 + 
\msq{\stopl} |\stopl|^2 + \msq{\stopr} |\stopr|^2 - \notag \\ 
&\; B_{\mu} \lb \hu \hd + {\rm c.c.} \rb  + \lb \yuk{t} A_{t} \hu \stopl \stopr +  {\rm c.c.} \rb - \lb \yuk{t} \mu \stopl \stopr \hd^{*} + {\rm c.c.} \rb + \notag \\
&\; \yuk{t}^2 \lb | \stopl \stopr |^2 + | \hu \stopl |^2 + | \hu \stopr |^2 \rb + 
\frac{g_2^2}{8} \lb |\hu|^2 - |\hd|^2 - |\stopl|^2 \rb^2 + \notag \\
&\; \frac{g_1^2}{8} \lb |\hu|^2 - |\hd|^2 + \frac{1}{3} |\stopl|^2 - \frac{4}{3} |\stopr|^2 \rb^2 + \frac{g_3^2}{6} \lb |\stopl|^2 - |\stopr|^2 \rb^2\ .
\label{scalar_potential}
\end{align}
The SML like minima exist in the 
$\stopl=\stopr=0$ hyperplane, as evident from the above
expression.  
Away from this plane, in the flat direction of quartic terms
($\stopl$=$\stopr$), quantities like 
$\yuk{t} A_{t}\hu\stopr\stopl$ and $\yuk{t} \mu \stopl \stopr \hd^{*}$ 
may become large and negative.  
For large values of $A_{t}$ and/or $\mu$ 
the above two terms may lead to global minima which 
break  $\rm U(1)_{EM},~SU(3)_C$ and the global 
U(1$)_{\rm Baryon}$ symmetries.  
Similar effects occur while \Vevs~are considered for 
$\tilde{b_R}$, $\tilde{b_L}$ or even for $\staul$ and $\staur$.  Hence
it is very important to probe vacuum stability for large 
$|\mu|$ and large A-parameters of third generation in the context of CCB.  
If the 
global minima is charge and color breaking, it is essential to evaluate the 
tunneling rate from SML to CCB minima for estimating 
the lifetime of the metastable SML state.  This lifetime will 
ultimately determine 
the viability of the corresponding MSSM 
parameter point\cite{Camargo-Molina:2013sta,Chowdhury:2013dka,Blinov:2013uda,Camargo-Molina:2014pwa}.

\noindent 
Semiclassical calculations of the false vacuum 
decay via quantum-tunneling through a barrier, may be performed for 
a single scalar field $\phi$(x) that resulted into the 
transition probability per unit time per unit volume as given below  
\cite{Kusenko_paper1,kusenko2,Brandenberger:1984cz}.  
\begin{align}
\Gamma/V = A\mathrm{e}^{-S[\bar\phi]/\hbar}.
\label{decay_eqn}
\end{align}
Here, $\bar\phi$ is 
a particular configuration of the field $\phi$ for which $\delta$S=0.  
This field configuration which dominates the integral 
is called a {\it bounce}\footnote{For details of semiclassical 
calculation of vacuum decay and related issues 
see \cite{coleman}.},~which 
is a stationary point of the Euclidean action.  The bounce is a  
non-trivial solution of the Euclidean
Euler-Lagrange equation that obeys specific boundary conditions.  
The probability for the Universe to have decayed to a deeper CCB minima 
by the 
present time $t_0$, the age of the Universe 
is roughly equal to $t_{0}^4\times\Gamma/V$.  Here $t_0^4$ refers to 
an estimate of a four-volume within which the transition may take 
place.  
Considering a 100 GeV scale, thereby ${\rm A}\simeq~(100~ \gev)^4$, 
one obtains $S_E[\bar\phi]/\hbar\sim 400$ for $t_{0}^4\times\Gamma/V\sim 1$.  
Therefore, the SML false vacuum at which the Universe rests at the present 
time may be considered to be stable against decay 
for $S_E[\bar\phi]/\hbar>400$\cite{Kusenko_paper1,kusenko2}, indicating a long-lived scenario.
At this point we stress the need of numerical computation. 
For the simplest case of a single scalar field, explicit 
analytic calculations may be performed under certain
approximations namely the {\it thin wall} 
and the {\it thick wall} 
scenarios~\cite{Kusenko_paper1,kusenko2,Brandenberger:1984cz}.  
On the other hand, an accurate analysis which involves multiple 
scalar fields may not divide itself into thin or thick wall zones 
for phenomenologically significant regions of parameter space.
Therefore, one must take resort to numerical computation
to determine the fate of SML vacuum in presence of deeper CCB vacua 
as performed in {\tt Vevacious}\cite{Camargo-Molina:2013qva}.  

\noindent 
Physics of CCB minima and phase transition in the Early Universe has 
important implication 
on the evolution of the 
Universe and its present ground state~\cite{Kusenko_paper1,kusenko2,Brandenberger:1984cz}.  
As par the previous discussion, the depth of the 
CCB minima depends on squark/slepton mass terms, 
the relevant trilinear coupling parameters and $\mu$.  
At a finite temperature the 
scalar potential is modified by terms $\propto$~$T^2$ which are 
similar to mass square terms.  The trilinear terms also receive 
corrections $\propto$~T. The history of the potential goes as follows\cite{Brandenberger:1984cz}.  
\begin{itemize}
\item At a very high temperature, the potential is symmetric with one 
minima at $\phi=0$. 
\item Afterwards, at a critical temperature $T_c$ degenerate 
minima occur for vanishing as well as non-vanishing $\phi$.
\item
As the temperature decreases with time, the degeneracy 
breaks and the minima at non-zero $\phi$
becomes deeper.  
\item
Finally, at T=0, there is a maxima at $\phi=0$ and minima
at some non-zero $\phi$ that corresponds to ordinary SML SSB ground state 
(for $\phi$ being a neutral colorless scalar). 
\end{itemize} 
Transition from an SML minima to a deeper CCB minima which 
is the focus of our discussion, is a first order phase transition\cite{Brandenberger:1984cz}.  

\noindent 
We shall now discuss the relevance of studying electroweak 
vacuum stability in the global CCB scenario, in the post Higgs@125~GeV era 
in which the present data of the Higgs boson mass is 
125.7$~\pm~0.6\rm{~\gev}$~\cite{HiggsDiscoveryJuly2012}.  
There is a high chance that the discovered Higgs boson is SM like\cite{Chatrchyan:2013lba}.  
Hence, in MSSM it would correspond to the CP-even lightest Higgs boson $h$ 
assuming a decoupling limit of Higgs scenario ($M_Z^2 \ll M_A^2$)\cite{djouadi}.
Certainly, with a tree level bound of  $M_Z^2~\cos^22\beta$ for $m_h^2$ 
one requires a large radiative corrections that on the other hand,  
push the super-partner spectra on the higher side in unified models. In MSSM  
this translates to the requirement of heavy top-squarks. 
The dominant loop correction that is due to the top-stop loops is 
given by\cite{djouadi} 
\begin{align}
\Delta m_{h,top}^2= \frac{3 g_2^2 {\bar m}_t^4}{8 \pi^2 \MW^2} \left[\ln\left(\frac{\ms{t}{1} \ms{t}{2}}{{\bar m}_t^2}\right) + \frac{X_t^2}{\ms{t}{1}\ms{t}{2}} 
\left(1 - \frac{X_t^2}{12\ms{t}{1}\ms{t}{2}} \right) \right]. 
\label{stop_loop}
\end{align}
Here $X_t=A_t-\mu\cot\beta$ and ${\bar m}_t$ stands for the running top-quark 
mass that includes electroweak, QCD and SUSY QCD 
corrections\cite{Pierce:1996zz}.~~    
With the requirement of the above large loop corrections in 
the post Higgs discovery scenario, one must explore the regions of 
parameter space that do not demand so high stop masses but the 
effect would come from the term involving $X_t$ 
in the above equation. The maximal mixing scenario of $X_t=\sqrt 6 M_S$, 
where $M_S=\sqrt{\ms{t}{1}\ms{t}{2}}$ is certainly useful
\cite{SUSYreviews2,SUSYbook1,SUSYbook2,djouadi}.  
However as we will see in the next section, scenarios of large $\mu$ are associated
with a significant amount of radiative corrections from sbottom and stau loops.  
As a result, maximal mixing may occur away from $\sqrt 6 M_S$ for $X_t$. We will come to this
point soon.
While exploring the MSSM parameter space, we must be careful that the MSSM scalar potential  
may have CCB minima deeper than the 
SML minima. Therefore, the study of stability of the SML false 
vacuum against decay to global CCB minima is extremely important. 
Quite naturally it becomes important to check the degree of effectiveness of the inequalities related to CCB 
constraints in this regard both for stable and long-lived vacuum states, 
as we will discuss below.  

\noindent 
Suitable analytic constraints were imposed on the relevant
MSSM parameters to avoid the appearance
of CCB global minima in Refs. \footnote{Traditional bounds were initially  
studied in Refs.~1~and~2 of Ref.\cite{Casas:1995pd}. For CCB 
constraints in models beyond MSSM see Ref.\cite{Ellwanger:1999bv}} 
\cite{AlvarezGaume:1983gj,Gunion:1987qv,Casas:1995pd,
Strumia:1996pr,Baer:1996jn,Brhlik:2001ni,LeMouel:2001ym,Bordner:1995fh}.  The nature of CCB 
minima and consequently the constraints 
depend on the particular nature of the \Vevs. In order to derive simpler 
analytical constraints, assumptions are made 
depending on the inter-relationships of \Vevs~ that isolate suitable directions 
in the field space.
In the direction ``b'' discussed in Ref.\cite{Casas:1995pd}, 
non-vanishing \Vevs~were considered for 
$|H_u|, |H_d|, |Q_u|, |u_R|$ as well as possibly for $|L_i|$ 
that under simplifying assumptions of D-flat directions 
lead to\footnote{Following usual notation of fields such as that of Ref.\cite{SUSYbook1}.}
\begin{align}
A_u^2\leqslant~3[m_2^2+m_{Q_u}^2+m_u^2].
\label{traditional_bound1}
\end{align}
Here $m_2^2=m_{H_u}^2+\mu^2$. The above bounds are 
imposed on all the three generations of up-type squarks and traditionally 
used in popular SUSY spectrum generators that also use conditions for avoiding 
potential to be unbounded from below 
\cite{ufbpapers}.  
On the other hand in the 
direction ``a'' of Ref.\cite{Casas:1995pd} the authors considered 
non-vanishing \Vevs~ of $|H_u|, |Q_u|, |u_R|$ as well as that of $|d_L|,|d_R|$ or possibly 
$|L_i|$, but vanishing \Vev~ of $|H_d|$ which under simplifying assumptions resulted into 
the following inequality.
\begin{align}
A_u^2\leqslant~3[m_2^2-\mu^2+m_{Q_u}^2+m_u^2].
\label{traditional_bound2}
\end{align}
Using $m_{H_u}^2\approx-\mu^2$, the above reduces to
\begin{align}
A_u^2+3\mu^2\leqslant~3[m_{Q_u}^2+m_u^2].
\label{traditional_bound2a}
\end{align}
Constraints of Eq.\ref{traditional_bound2a} are valid for small 
Yukawa couplings. However, if one uses the above for top Yukawa coupling  
one effectively obtains a stronger bound, embracing the traditional 
bounds of Eq.\ref{traditional_bound1} \cite{Casas:1995pd}. Going beyond the exact CCB 
constraints, the possibility of existence of 
long-lived SML minima were considered afterwards. 
Thus Ref.\cite{Kusenko_paper1} and recently Ref.\cite{Cohen:2013kna} 
incorporated the 
above long-lived scenario to come up with the following inequalities that 
obviously allowed an enlarged parameter space,  
\begin{equation}
A_u^2+3\mu^2 \leqslant  7.5[m_{Q_u}^2+m_u^2], 
\label{relaxed_bound1}
\end{equation}
\begin{equation}
A_u^2 \leqslant  ~3[m_{H_u}^2+\mu^2]+7.5[m_{Q_u}^2+m_u^2]. 
\label{relaxed_bound2}
\end{equation}
We now try to discuss under what conditions the analytic constraints 
were evaluated. 
The simple bounds of Eq.\ref{traditional_bound1} were obtained considering the 
D-flat directions assuming \Vevs~ of the concerned scalar fields to be equal. 
Analysis of realistic scenarios must involve unequal \Vevs~ that is also 
sufficiently case specific\cite{Casas:1995pd}.  
However, as we will discuss below, 
even when simplistic 
assumptions are made for D-flat directions 
it is seen that more robust bounds such as 
Eq.\ref{traditional_bound2a} compared to Eq.\ref{traditional_bound1} may lead 
to unnecessary degree of stringency\cite{Chowdhury:2013dka}. The same is true for 
the long-lived scenarios of Eq.\ref{relaxed_bound1} and 
Eq.\ref{relaxed_bound2} which are found to be neither necessary nor 
sufficient\cite{Chowdhury:2013dka,Blinov:2013uda,Camargo-Molina:2014pwa}.
As mentioned previously in this work similar to the 
Refs.\cite{Carena:2012mw1,Kitahara:2013lfa,Camargo-Molina:2013sta,
Chowdhury:2013dka,Blinov:2013uda,Camargo-Molina:2014pwa} we will follow the 
numerical route to analyze the CCB constraints while considering 
the possibility of existence of long-lived SML minima.  
We have used {\tt Vevacious}\cite{Camargo-Molina:2013qva}~ for our 
analysis. Combined with a SUSY spectrum generator 
the code finds the
global minima of the associated scalar potential.  
In absence of global 
CCB minima the SML vacuum is stable. If the global
minima is found to break the charge and color symmetry, the code computes 
the lifetime of the SML minima against decay to the global CCB minima using 
the code {\tt CosmoTransitions}\cite{Wainwright:2011kj}. It then determines 
whether the SML minima is long-lived or short-lived.
We will particularly probe the pMSSM parameter space in detail for 
large $|\mu|$ and large $|A_t|$ zones while keeping any issue 
related to naturalness\cite{naturalness,naturalness_recent,naturalness_others} aside.   
Large $|\mu|$ is often considered in works involving global 
analyses of parameter space for unified models like 
CMSSM\cite{Roszkowski:2014wqa}. It has also been explored in pMSSM 
related studies\cite{Cahill-Rowley:2014boa} or a similar 
analysis involving vacuum stability as in 
Ref.\cite{Bobrowski:2014dla}. 

\section{Results}
Here we present the results of our analysis over different
regions of parameter space of pMSSM and 
classify parameter ranges according to stable, long-lived or short-lived 
vacua in different subsections. First, 
we discuss vacuum stability in the generic 
region of parameter space, in particular for low values of $\mu$.  Later, 
we extend our analysis for a large value of $\mu$ while 
varying both stop and sbottom sector parameters and assigning \Vevs~to stop 
and sbottom scalar fields.  Then we scan over a wide range of values of  
$\mu$ and $A_t$ for a moderate and a large value of $\tan\beta$ 
exploring interesting regions 
of pMSSM parameter space that may potentially fall in the zone of 
maximal radiative corrections to Higgs boson. 
This section also analyses the impact of dark matter namely neutralino relic density 
and direct detection limits on our work.  Additionally we will discuss 
the compatibility of our analysis with 
muon $g-2$ data.  The relevant SM parameters used are $m_t^{pole}= 173.5~\gev$ ,
$m_b^{\overline{MS}} = 4.18~\gev$ and $m_{\tau}=1.77~\gev$.
\subsection{Study of generic region of pMSSM parameter space for the stability of vacuum}
\label{generic mssm}
\begin{figure}[htb]
\begin{center}
\includegraphics[height=75mm]{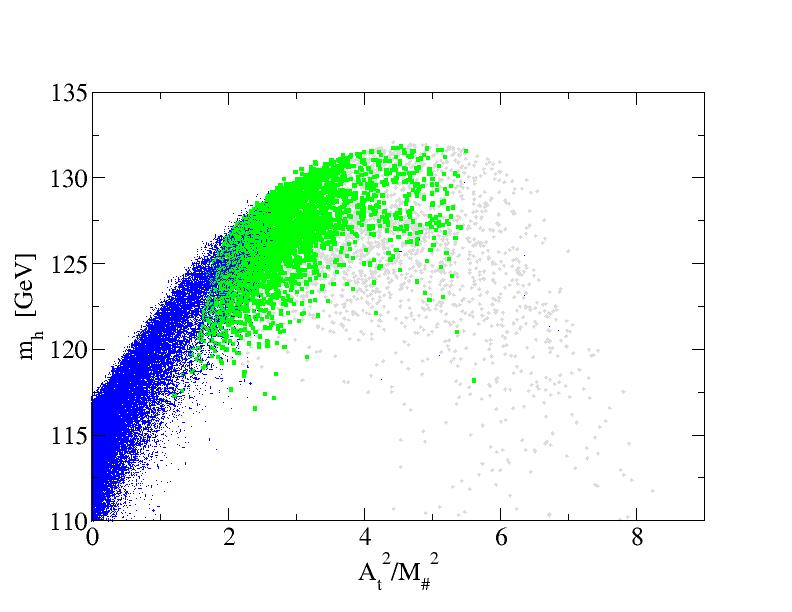}
\caption[]{\it The variation of $m_h$ against $A_t^2$/$M_\#^2$ for the 
scanning ranges of Eq.\ref{scanmssm1}.  
Blue,~green,~grey dots 
correspond to stable, long-lived and short-lived vacua respectively. The first two type will comprise ``safe'' vacuum.}
\vskip 0.3cm
\label{fig:generic_MSSM_1}
\end{center}
\end{figure}

In this part we analyze the CCB constraints by focusing on 
generic part of pMSSM parameter space, in particular for low values 
of $\mu$ and non-vanishing $A_t$ that is important for the 
Higgs mass limit.  Here the parameter space 
spans a broad range of $\tan\beta$, third generation 
of up-type squark masses, $A_t$ and $\mu$ upto a \tev. 
In this subsection, in order to compare our results with 
Refs.\cite{Chowdhury:2013dka,Blinov:2013uda,Camargo-Molina:2014pwa} 
we allow only stop fields ($\stopl$ and $\stopr$) to take non-zero \Vevs ~along with the 
Higgs fields.
Our choice of parameters are as follows. 
\begin{eqnarray}
\label{scanmssm1}
500~\gev&\leqslant m_{\tilde{Q}_{3}}~&\leqslant~1500~\gev, \notag \\
500~\gev&\leqslant~m_{\tilde{U}_{3}}~&\leqslant~1500~\gev, \notag \\
5 & \leqslant~\tan\beta~&\leqslant~60,\\
100~\gev&\leqslant~\mu~&\leqslant1000~\gev,\notag \\
-3~m_{\tilde{Q}_{3}}&\leqslant A_t&\leqslant 3~m_{\tilde{Q}_{3}}.\notag 
\end{eqnarray}
We set all other sfermion masses to be at 1 \tev,~$\rm M_A$= 1 \tev.  The gaugino masses are fixed at 
$\rm M_1 = 100$ ~\gev,~$\rm M_2  = 300$ $~\gev$~and $\rm M_3 = 1000~\gev$.
All other trilinear couplings are set to zero.
The other sfermion masses or 
the gaugino masses could indeed be chosen at different zones of values without 
essentially affecting the results of our analysis involving long-lived states. \\Keeping an eye on 
Eq.\ref{traditional_bound1} we define 
$M_\#^2=m^2_{H_2}+\mu^2+\msq{\stopl} + \msq{\stopr}$ and plot 
Fig.~\ref{fig:generic_MSSM_1} for variation of 
the lightest Higgs boson mass $m_h$ vs the 
 dimensionless quantity $A_t^2/M_\#^2$. 
This would easily identify our results  
with respect to the traditional bounds of CCB constraint and additionally 
show the validity zones of the long-lived states.
Blue, green and grey colored points correspond to stable, long-lived 
and short-lived vacuum states respectively. It is evident from the plot that there
exist safe vacua (long-lived and stable states) where the traditional constraint 
(Eq.\ref{traditional_bound1}) is violated.  
There is a significant 
zone of long-lived vacuum states where the Higgs mass is quite high, close to  
even the maximum value (see Eq.\ref{higgsrange} and related discussion as given below) 
a fact generally 
consistent with the results of 
Refs.\cite{Chowdhury:2013dka,Blinov:2013uda,Camargo-Molina:2014pwa}.  
This shows the 
importance of considering the existence of long-lived states in the post 
Higgs@125~\gev~era. 
Additionally, even if we do not consider the long-lived 
states, our numerical exploration of CCB constraints shows that 
Eq.\ref{traditional_bound1} is only approximately valid, for example 
there exist green regions below $A_t^2/M_\#^2=3$. 

\begin{figure}[H]
\begin{center}
\includegraphics[height=65mm]{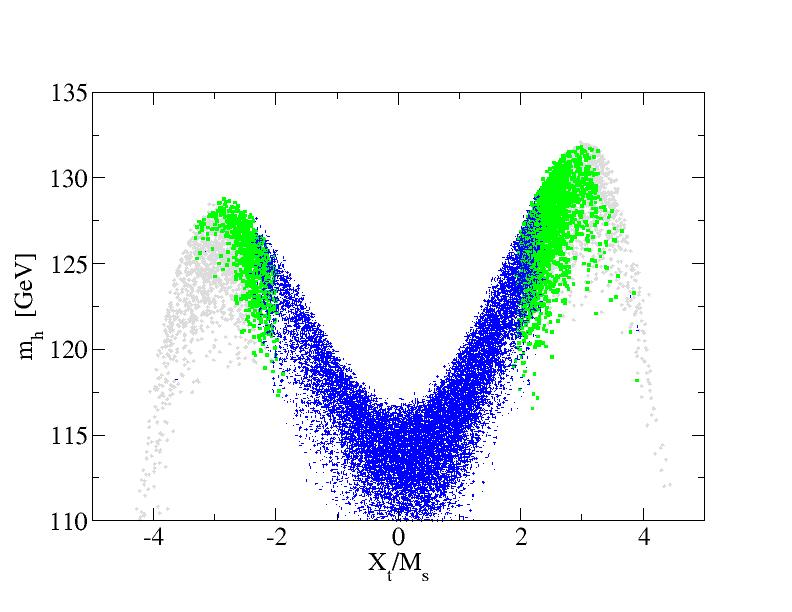}
\caption[]{\it The variation of $m_h$ vs $X_t/M_S$ for the scanning ranges of \ref{scanmssm1}.
Blue, green and grey dots corresponds 
to stable, long-lived and short-lived vacua respectively.  }
\label{fig:generic_MSSM_2}
\end{center}
\end{figure}

\noindent 
A similar result when projected into $m_h$ vs $X_t/M_S$ plane, 
(where $M_S=\sqrt{m_{\stopl} m_{\stopr}}$) appears in 
Fig.~\ref{fig:generic_MSSM_2}. Here $m_h$ maximizes itself 
in both the regions, negative and positive for $X_t$, where there exist 
long-lived rather than stable states. 
We like to mention here that 
considering the existing uncertainties in the computation of radiative 
corrections to Higgs mass 
we assume a 3~\gev~ window in $m_h$ leading to the  
following range\cite{loopcorrection}. This could arise from renormalization scheme 
related uncertainties, scale dependence, problems in computing higher 
order loop corrections up to three loops or the uncertainty 
in the experimental value of top-quark mass 
\footnote{We also remind the reader the additional issue of uncertainty 
of about 2.8 GeV in $m_t^{pole}$ as argued in Ref.\cite{Alekhin:2012py,Chakraborti:2012up}.}.
\begin{equation}
122\leqslant~m_h~\leqslant~128~\gev.
\label{higgsrange}
\end{equation} 

\noindent 
We would like to mention that we have imposed the constraints from 
${\rm Br}(B \rightarrow X_s \gamma)$ as well as 
${\rm Br}(B_s \rightarrow \mu^+ \mu^-)$ in this analysis and the shown 
parameter points completely satisfy the following conditions irrespective of the 
nature of the vacuum. 
The experimental limits on
${\rm Br}(B \rightarrow X_s \gamma)$ as given in~\cite{Amhis:2012bh} is ~
${\rm Br}(B \rightarrow X_s \gamma)=[3.42\pm0.22]\times 10^{-4}$ which at $3\sigma$ level results into
\begin{align}
2.77\times 10^{-4}~\leqslant~{\rm Br}(B \rightarrow X_s \gamma)~\leqslant~4.09\times 10^{-4}.
\label{bphysics_constraints1}
\end{align}
The recent constraints from~${\rm Br}(B_s \rightarrow \mu^+ \mu^-)$ as 
obtained from CMS and LHCb
~\cite{Aaij:2013aka,Chatrchyan:2013bka,CMSandLHCbCollaborations:2013pla} indicate ${\rm Br}(B_s \rightarrow \mu^+ \mu^-)=[2.9\pm 0.7]\times 10^{-9}$, 
which at $3\sigma$ level leads to 
\begin{align}
0.8 \times 10^{-9}~\leqslant~{\rm Br}(B_s \rightarrow \mu^+ \mu^-)~\leqslant~5\times 10^{-9}. 
\label{bphysics_constraints2}
\end{align}
We compute the above B-Physics results using {\tt SuperIso}
\cite{Mahmoudi:2007vz} while using {\tt SuSpect}\cite{Djouadi:2002ze} as the 
spectrum generator.
  In the following subsection we go 
beyond the generic region of pMSSM parameter space of 
Refs.\cite{Chowdhury:2013dka,Blinov:2013uda,Camargo-Molina:2014pwa}, 
and explore the role of large $|\mu|$ and large $|A_t|$ in the 
context of the discussion made in Sec.\ref{ccbmssm}.
\vskip 0.5cm
\subsection{Stability of vacuum for a fixed $\tan\beta$ and a large $|\mu|$}
\label{large_mu_scenario}
In order to analyze with large values of $|\mu|$ we must remember that radiative 
corrections 
to Higgs boson mass from sbottom and to a lesser degree from stau sectors may hardly be 
ignored. This is because of a quartic dependence on $|\mu|\tan\beta$
\cite{Altmannshofer:2012ks,Bhattacherjee:2013vga} which is albeit suppressed by a quartic dependence  
of scalar masses, in addition to the effect of smallness of $m_b$ or $m_\tau$ compared to $m_t$. 
Similar to Eq.\ref{stop_loop} the corrections from sbottom sector read\cite{Altmannshofer:2012ks,Bhattacherjee:2013vga},
\begin{align}
\label{sbottomeqn}
\Delta m_{h,bottom}^2= \frac{3 g_2^2 {\bar m}_b^4}{8 \pi^2 \MW^2} \left[\ln\left(\frac{\ms{b}{1} \ms{b}{2}}{{\bar m}_b^2}\right) + \frac{X_b^2}{\ms{b}{1}\ms{b}{2}} 
\left(1 - \frac{X_b^2}{12\ms{b}{1}\ms{b}{2}} \right) \right],
\end{align}
where $X_b=A_b-\mu \tan\beta$\footnote{A similar result for the stau
contribution would involve $X_\tau=A_\tau-\mu\tan\beta$.}. 
Henceforth we allow sbottom fields in addition to stop fields 
to acquire non-zero \Vevs~besides the Higgs fields within {\tt Vevacious}.
We fix $\mu$ at 9~\tev,~$\rm M_A$ at 1~\tev~ and 
$\tan\beta=20$. Our range of scanning as given below 
involves parameters related to the stop and the sbottom sectors. 
\begin{align}
\label{scan-ranges}
500~\leqslant~m_{\tilde{Q}_{3}}\leqslant~3000\ \gev, \notag \\
500~\leqslant~m_{\tilde{U}_{3}}\leqslant~3000\ \gev, \notag\\
500~\leqslant~m_{\tilde{D}_{3}}\leqslant~3000\ \gev, \\
-10~\leqslant~A_t~\leqslant~10\ \tev, \notag\\
-10~\leqslant~A_b~\leqslant~10\ \tev. \notag
\end{align}
We keep all other scalar mass parameters fixed at $1~\tev$.  We focus on a pocket of pMSSM parameter space as a
representative zone that satisfy the cold dark matter 
constraints from WMAP\cite{Hinshaw:2012aka}/PLANCK\cite{Ade:2013zuv}.  
This corresponds to the following 
gaugino mass parameters  
\begin{align}
\rm M_1=500~\gev,~
\rm M_2=525~\gev,~
\rm M_3=1400~\gev~.
\end{align}
Our choice of $M_3$ is consistent with the recent limits on 
$m_{\sg}$\cite{Aad:2014wea}. 
The generic overabundance of a bino-like 
lightest neutralino dark matter which is the 
lightest supersymmetric particle (LSP) is brought under control 
via appropriate bino-wino ($\widetilde{B}-\widetilde{W}$) 
coannihilation\cite{Edsjo:1997bg,Gondolo:1990dk}.  
This choice also satisfies the LUX\cite{Akerib:2013tjd} 
limit for the spin-independent direct detection cross section 
$\sigmasip$. Both the relic density and $\sigmasip$ are computed using 
{\tt micrOMEGAs}\cite{micromegas_ref}.
\begin{figure}[H]
\begin{center}
\includegraphics[height=70mm]{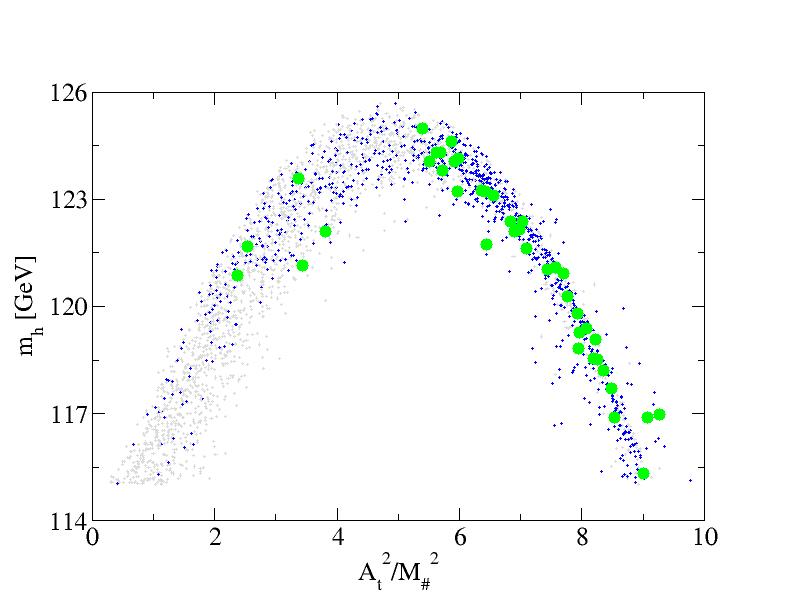}
\caption[]{\it Plot of $m_h$ vs $A_t^2/M_\#^2$ corresponding to 
the scan of Eq.\ref{scan-ranges} for $\mu=9$~TeV.  
Green, blue and grey dots correspond to
long-lived, stable and short-lived SML vacuum 
states respectively.  
The points are spread throughout the plane without much 
clustering effect unlike Fig.\ref{fig:generic_MSSM_1}. Lack of clustering and appearance 
of long-lived states in the right half of the figure clearly shows the 
absence of validity of Eqs.\ref{traditional_bound1} and \ref{traditional_bound2a} when $\mu$ is considered appreciably 
large. 
}
\label{fig:large_mu_1}
\end{center}
\end{figure}

\noindent 
In Fig.~\ref{fig:large_mu_1} we show the 
variation of $m_h$ against $A_t^2/M_\#^2$ corresponding to 
the scan of Eq.\ref{scan-ranges} for $\mu=9$~TeV.  
Green, blue and grey dots that satisfy all the experimental limits 
mentioned above correspond to
long-lived, stable and short-lived SML like vacuum 
respectively.
Unlike Fig.\ref{fig:generic_MSSM_1}, the points are 
spread throughout the plane without much 
clustering effect. Lack of clustering and appearance 
of long-lived vacuum in the right half of the figure 
clearly shows the absence of validity of Eqs.\ref{traditional_bound1} and \ref{traditional_bound2a} when 
$\mu$ is considered appreciably large.
\subsection{Scan over wide range of $\mu$ and $A_t$ for $\tan\beta=20$} 
\label{largemuscansection}

\noindent 
In order to probe the impact of $\mu$ and $A_t$
on the stability of vacuum we scan over the same parameters in a 
wide range for a given set of pMSSM input values.  
Considering $\tan\beta=20$, we fix $m_{\tilde{Q}_{3}}, m_{\tilde{U}_{3}},
 m_{\tilde{D}_{3}}$~ at 2~$\tev$.    All other sfermion masses are
fixed at 1~$\tev$ and $M_A$ is taken to be 1~$\tev$.  
While allowing $\stopl$,$\stopr$,$\sbottoml$~and
$\sbottomr$~ to acquire non-zero \Vevs~along with the Higgs 
fields we choose the following range for $\mu$ and $A_t$.    
\begin{eqnarray}
\label{mu_at_scan}
%\centering
-10~\tev~\leqslant~A_t~&\leqslant&~10~\tev,\ \notag\\
-11~\tev~\leqslant~\mu~&\leqslant &~11~\tev\  . 
\end{eqnarray}
We should keep in mind the importance of non-vanishing
$A_b$ in context of vacuum stability in CCB scenario,
particularly for large $\mu$ zones away from the generic region. However, 
non-vanishing $A_b$ would hardly have an effect on $m_h$.
Hence we use the following range for $A_b$ namely, $-6~\tev $ to $6~\tev$ 
and consider vanishing trilinear couplings except $A_t$. 
Similar to Sec.\ref{large_mu_scenario} we choose the same 
gaugino mass parameters,  
whereas we impose B-Physics constraints of 
Eqs.~\ref{bphysics_constraints1} and \ref{bphysics_constraints2} on the 
resulting spectrum as before.  

\begin{figure}[htb]
\begin{center}
\includegraphics[height=90mm]{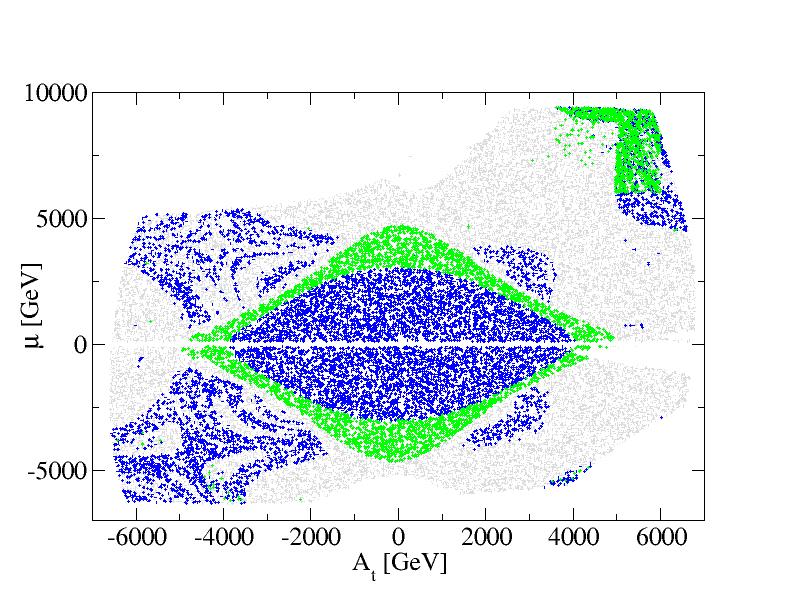}
\caption[]{\it Plot of $\mu$ vs $A_t$ for $\tan\beta=20$ with fixed pMSSM parameters described in 
the beginning of Sec.\ref{largemuscansection}.  
Green, blue, grey dots corresponds to
long-lived, stable and short-lived SML vacuum states.  The central zone of
stable states and the surrounding strip of long-lived states include the generic region 
of pMSSM parameter space, that is characterized by comparatively lower value of $\mu$. Interestingly
there exist pockets of safe vacuum states in the zone much away from the central region, where traditional CCB
constraints of Eqs.\ref{traditional_bound1}~or~\ref{traditional_bound2a} are violated.}
\label{fig:mu_scan_1}
\end{center}
\end{figure}

\noindent 
In Fig.~\ref{fig:mu_scan_1}~ we show the distribution 
corresponding to safe and dangerous SML
vacua in the $\mu-A_t$~plane.  The central blue zone denotes the stable 
vacuum and the surrounding green
strip represents the long-lived SML vacuum states.  
It is evident that the above zone that also includes small values of $\mu$, 
referred in this analysis as generic zone, is 
symmetric over both $\mu$~and~$A_t$.  
This implies that in the central zone, 
the stability of SML vacuum against decay to deeper CCB states is 
largely independent of the 
sign of $\mu$~and~$A_t$.  For larger $|\mu|$ within the central 
blue region, we see that safe vacua occur 
for smaller value of $|A_t|$
~and vice-versa.  

\noindent
Surrounding the central blue zone and the associated 
green peripheral region, one finds large grey regions designating   
short-lived vacuum states in which there also exist pockets of 
stable and long-lived zones in all the four quadrants. 
We further see the existence of stable/long-lived states 
for very large positive values of  $\mu$~and~$A_t$, which would however 
be excluded by the traditional analytic constraints.
Thus in Fig.~\ref{fig:mu_scan_1} where we have used constraints 
from B-physics but not the Higgs mass bound, we find that 
SML minima is extended to island areas with appreciably large 
positive values of both $\mu$~and~$A_t$.  We will later show that 
the island region of long-lived states with large $\mu$ and $A_t$ 
may also satisfy the Higgs mass limits. We particularly focus on  
the long-lived states\footnote{We have discarded the parameter 
points that typically give warning messages related to appearance of 
saddle points.} for the island region in the first quadrant and ignore 
any conclusion on the stable states in the same zone.  
This is connected to the fact that declaring a 
parameter point to be stable in the island region 
characterized by large $\mu$ may be quite non-trivial due to 
various computational issues\footnote{Private communication 
with the authors of {\tt Vevacious}.}.

\vskip 0.3cm
\subsubsection{Maximized $m_h$ zones in relation to long-lived states: 
Regions I and II} 
\begin{figure}[htb]
     \begin{center}
        \subfigure[]{%
            \label{fig:mh_mu}
            \includegraphics[width=0.51\textwidth]{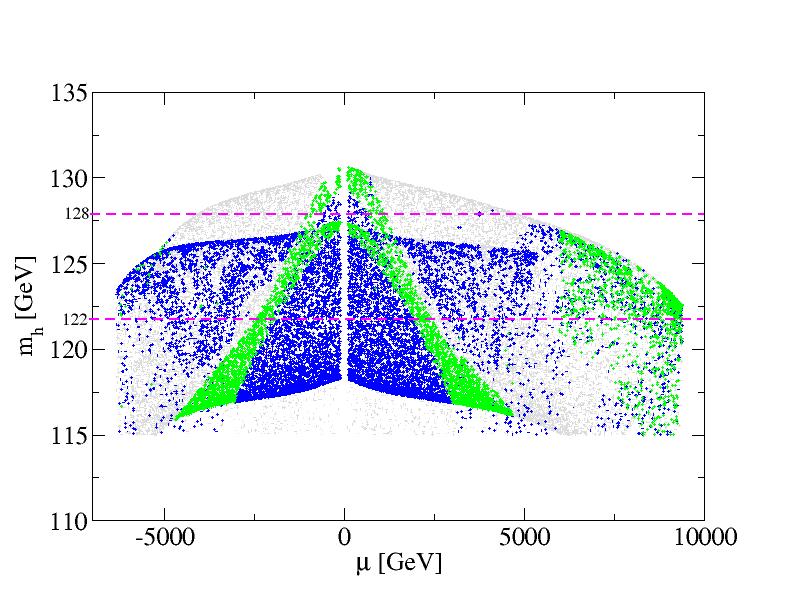}
        }%
        \subfigure[]{%
           \label{fig:mh_at}
           \includegraphics[width=0.51\textwidth]{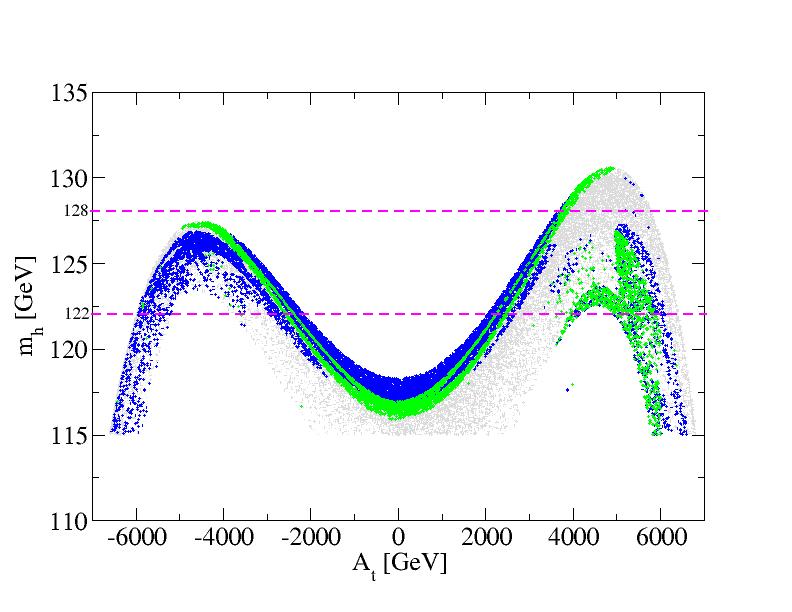}
        }
    \end{center}
    \caption{%
\it Fig.\ref{fig:mh_mu} shows the result of scanning described at the beginning of 
over $A_t$ in the 
plane of $m_h-\mu$ for $\tan\beta=20$.    
        The color codes for different vacuum stability 
conditions are same as that of Fig:~\ref{fig:mu_scan_1}.
        This confirms the existence of stable and long-lived 
vacua (satisfying 
        Higgs mass limits) in the region characterized by large values of 
$\mu$.
Fig.\ref{fig:mh_at} shows the result of scanning over $\mu$ in the 
plane of $m_h-A_t$.   
It turns out that there can be stable and long-lived states for large $|\mu|$ and/or 
large $|A_t|$ that would not satisfy the traditional CCB constraints of 
Eqs.\ref{traditional_bound1}~or~\ref{traditional_bound2a}. See text 
for Region-I and Region-II in relation to this figure.
     }%
   \label{fig:mh_mu_at}
\end{figure}

In Fig.~\ref{fig:mh_mu_at} we show the effect of the same 
scanning (Eq.\ref{mu_at_scan}) on Higgs boson mass $m_h$. 
In Fig.~\ref{fig:mh_mu} we show the blue, green and grey regions   
in $m_h$-$\mu$ plane.  
On finds symmetric distribution about the $\mu$-axis within the large 
triangular green area.  
Short-lived vacuum states denoted by grey dots occupy the region outside the 
green strip of long-lived states.  

\noindent
We now identify the parameter region in $\mu-A_t$ plane in 
relation to where $m_h$ maximizes {\rm i.e.} becomes 
close to the upper edge of 
Eq.\ref{higgsrange} as far as possible.   
We further check whether the maximized zones satisfy 
the traditional CCB constraints of Eqs.\ref{traditional_bound1}, \ref{traditional_bound2a} or whether
they fall into the category of long-lived vacuum states.
Hence, for the above purpose, staying within the valid band of 
$m_h$ (Eq.\ref{higgsrange}) 
we would particularly like to focus on two regions in Fig.~\ref{fig:mh_mu}. 
Region-I (long-lived) is identified with $|\mu| \simeq 1$~TeV 
and $m_h \sim 128$~GeV (we would call this as small $\mu$ zone), 
whereas Region-II (long-lived) 
occurs with $7 \lsim \mu \lsim 9$~TeV and $123 \lsim m_h \lsim 125$~GeV 
(we would refer it as the large $\mu$ zone). 
For Fig.~\ref{fig:mh_at} the same regions of Fig.\ref{fig:mh_mu} namely, Region-I maps to  
$A_t \sim \pm 4$~TeV, whereas Region-II is characterized by 
$3 \lsim A_t \lsim 5$~TeV.
Both Regions I and II are generally ruled out by at 
least one of the traditional CCB constraints of 
Eqs.~\ref{traditional_bound1} and \ref{traditional_bound2a} but they 
correspond to long-lived vacuum states. 
We further point out that although $m_h$ maximizes in Region-II for a 
given value of  
$\mu$, the corresponding value of $X_t$ appreciably differs from  
$\sqrt 6 {M_S}$ associated with an $m_h^{max}$ scenario\cite{djouadi}. 
This is indeed related to the discussion 
made regarding the radiative corrections to $m_h$ 
in Sec-\ref{ccbmssm}~specific 
to large values of $\mu$. 
Quite expectedly one finds that the relation  $X_t \simeq \sqrt 6 {M_S}$ 
holds good for low $|\mu|$ belonging to Region-I. 
In Fig.~\ref{fig:mh_at} similar to 
Fig.~\ref{fig:generic_MSSM_2}, $m_h$ maximizes 
for large positive values of $A_t$.  
Located symmetrically opposite to 
positive values of $A_t$, there is also a maximum of Higgs mass in the 
negative $A_t$ region.  This corresponds to 
a comparatively smaller value of $m_h$ in a long-lived vacuum scenario.  
\begin{figure}[htb]
\begin{center}
\includegraphics[height=90mm]{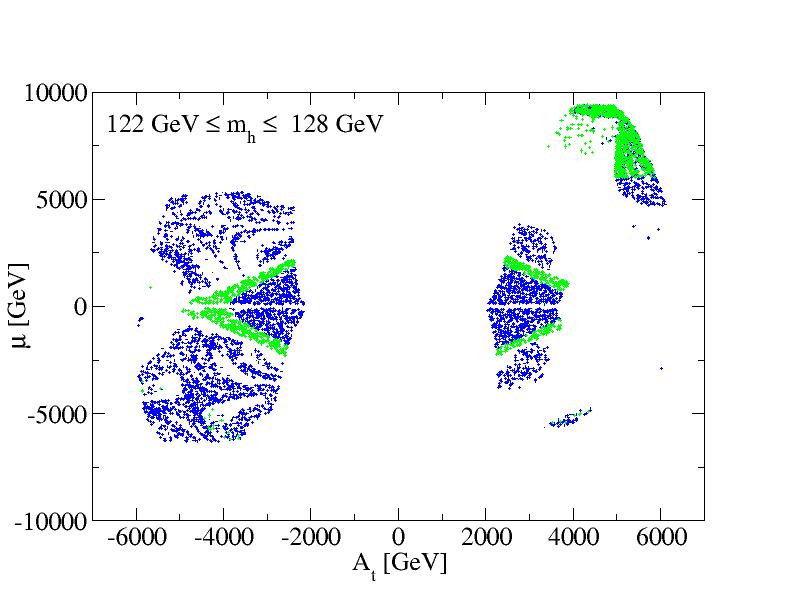}
\caption[]{\it Plot of $\mu$ vs $A_t$ for $\tan\beta=20$ after imposing limits on $m_h$ from Eq.\ref{higgsrange}.   
Blue and green dots represent stable and long-lived vacuum 
states respectively. 
It turns out there is a significant region of long-lived states much
away from the central region, where the 
traditional CCB constraints of 
Eqs.\ref{traditional_bound1}~or~\ref{traditional_bound2a} are violated.
The requirement to satisfy the limits of $m_h$ eliminates the small to moderate $|A_t|$ 
zones and the regions with small $|\mu|$ and very large $A_t$ due to 
radiative corrections to the Higgs boson mass. }
\label{fig:mu_4}
\end{center}
\end{figure}

\noindent 
Fig.\ref{fig:mu_4} is similar to 
Fig.~\ref{fig:mu_scan_1} but here we impose the limits on 
$m_h$ from Eq.\ref{higgsrange} and consider only 
the safe vacua in the $\mu-A_t$~plane. As expected, there is no valid 
region below $|A_t|=2$~TeV. 
There exists a significant area with large $\mu$ and large $A_t$ 
with stable (blue dots) and long-lived states (green dots) that satisfy 
the Higgs mass data. As mentioned before, most of the above regions 
on the other hand, would be excluded by the 
traditionally used CCB constraints of 
Eq.~\ref{traditional_bound1}~or~\ref{traditional_bound2a}. 
Thus the pMSSM parameter space can safely be extended to the above zone of   
large $|\mu|$ and large $|A_t|$. 
Since the latter zone corresponds to Region-II of Fig.\ref{fig:mh_mu} 
we infer that 
a maximized $m_h$ occurring in a region away from $X_t\simeq \sqrt 6 M_S$ 
would 
certainly require relatively smaller top-squark masses for a 
given amount of radiative corrections to the Higgs mass(see Eq.\ref{stop_loop}).  
We note that the requirement 
to satisfy the limits of $m_h$ eliminates i) small to moderate  $|A_t|$ 
zones and ii) regions with small $|\mu|$ and very large $A_t$.
\subsubsection{Compatibility with dark matter related constraints} 
\begin{figure}[htb]
\begin{center}
\includegraphics[height=90mm]{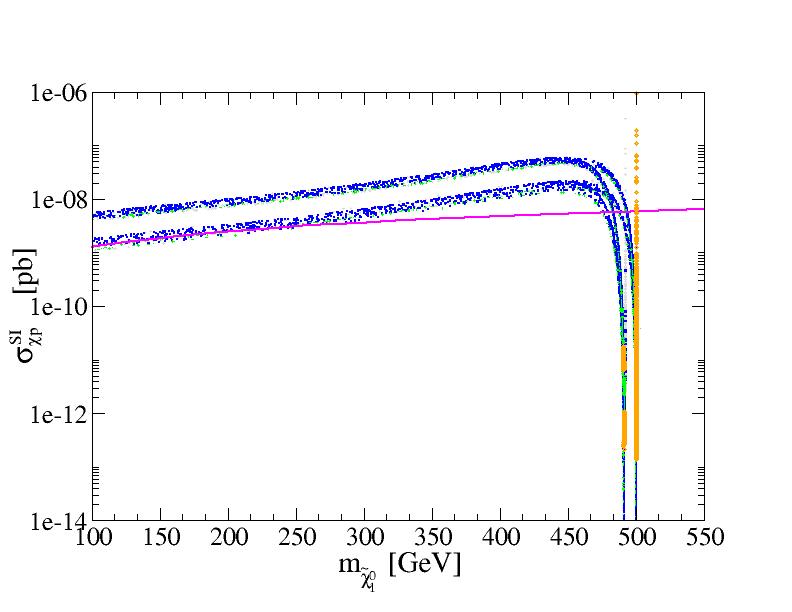}
    \caption{\it   
Plot of $\sigmasip$ vs $\mlspone$ for the scan of Sec.\ref{largemuscansection}.        
The region 
        above the pink line is excluded by the LUX limits on $\sigmasip$ \cite{Akerib:2013tjd}. 
        For $|\mu|< M_1$, there is a significant Higgino content in $\lspone$. Consequently
        $\sigmasip$ is large and $\varOmega_{\lspone}h^2$ is low. The presence of two 
        branches is attributed to slight dependence of $\sigmasip$ on the sign of $\mu$ that we
        have varied during the analysis.
        The orange colored region represents the zone 
        with proper relic abundance as mentioned in Eq.~\ref{omega_eqn}.  In these regions with adequate 
        dark matter abundance and allowed $\sigmasip$, $\lspone$ 
        is $\widetilde{B}$ dominated. Adequate relic abundance is 
        obtained via $\widetilde{B}-\widetilde{W}$ coannihilation. Most of the points characterized
        by small $\sigmasip$ cluster around two values of $\mlspone$ separated by a small amount. This 
        is due to the dependence of the radiative corrections to $\mlspone$ on the sign of $\mu$ as well as 
        on $A_t$ (via stop mass).
     }%
\label{fig:sigma}
\end{center}
\end{figure}

We now briefly discuss the compatibility of our parameter space with 
dark matter related data such as the relic density limits from 
WMAP\cite{Hinshaw:2012aka}/PLANCK\cite{Ade:2013zuv} and 
spin-independent direct detection $\lspone-p$ cross-section measurement from 
LUX~\cite{Akerib:2013tjd}. 
\\Figs.\ref{fig:sigma} shows the scatter plot 
of $\sigmasip$ vs $\mlspone$ with usual color codes.
Only a few points within the thin vertical lines near $\mlspone \simeq 
500$~\gev 
~satisfy the relic density ($\varOmega_{\lspone}h^2$) limits, shown in 
orange.  
  The imposed limits shown below at the level of 
$5\sigma$ of PLANCK~\cite{Ade:2013zuv} data 
accommodates well the range given by WMAP~\cite{Hinshaw:2012aka}.
\begin{align}
0.092\leqslant\varOmega_{\lspone}h^2\leqslant~0.138.
\label{omega_eqn}
\end{align}
We note that scattered points are clustered around two thin lines. It turns out that 
consideration of both signs of $\mu$ and $A_t$ leads to the appearance of 
two closely spaced lines. This is concerned with the slight dependence of the 
the radiative corrections to the mass of LSP\cite{Pierce:1993gj,Lahanas:1993ib,Akcay:2012db} 
on the signs of $\mu$, $A_t$ and the dependence of $\sigmasip$ on the sign of 
$\mu$\cite{Grothaus:2012js}.
\vskip 0.3cm
We emphasize here that giving \Vevs~ to several scalar fields within 
{\tt Vevacious} demands a large increase in computational time 
of the analysis. 
Hence, we could not afford to scan the gaugino mass parameters 
$M_1$ and $M_2$ that have immediate effects on $\mlspone$.   
Thus, we only probe the acceptability of the chosen point 
in the $M_1-M_2$ plane while we scan the specific pMSSM parameters 
relevant to CCB constraints. 
Our analysis  
involves a wide variation over the value of $\mu$. Hence,  
there are only a few points where one has $|\mu|<M_1$ for the chosen value $M_1=500$~\gev.   
These are 
the parameter points where the relic density is very small because 
of a strong presence of Higgsino within the LSP. At the same time,  
in the region of $\mu$ not far from $M_1$ or $M_2$, 
there can be a large bino-higgsino or even 
bino-wino-higgsino mixing which leads to a larger value of  
$\sigmasip$\cite{Hisano:2004pv}. 
This is confirmed in Fig.~\ref{fig:sigma} that shows larger 
$\sigmasip$ for $\mlspone$ below 480~\tev ~or so, mostly exceeding the 
LUX data~\cite{Akerib:2013tjd}. 
On the other hand, for larger values of $|\mu|$ when the LSP becomes almost  
a bino, we expect $\sigmasip$ to be small, a fact confirmed by the figure.  
A part of the above region characterized by small $\sigmasip$ provides 
correct relic density via bino-wino 
coannihilation\cite{Hisano:2004pv,Baer:2005jq}.
We comment that our chosen values of $M_1$ and $M_2$ 
that is consistent with  WMAP/PLANCK 
data would only be probed in future experiments 
like XENON1T\cite{Aprile:2012zx}. We like to point out that our analysis could be 
carried out for other appropriate gaugino masses that would satisfy the relic 
density and would result into $\sigmasip$ in the vicinity of the 
sensitivity region of LUX or future XENON1T experiments.
\subsection{Scan over wide range of $\mu$ and $A_t$ for 
$\tan\beta=40$ } 
\label{largetanbeta}
\begin{figure}[H]
\begin{center}
\includegraphics[height=90mm]{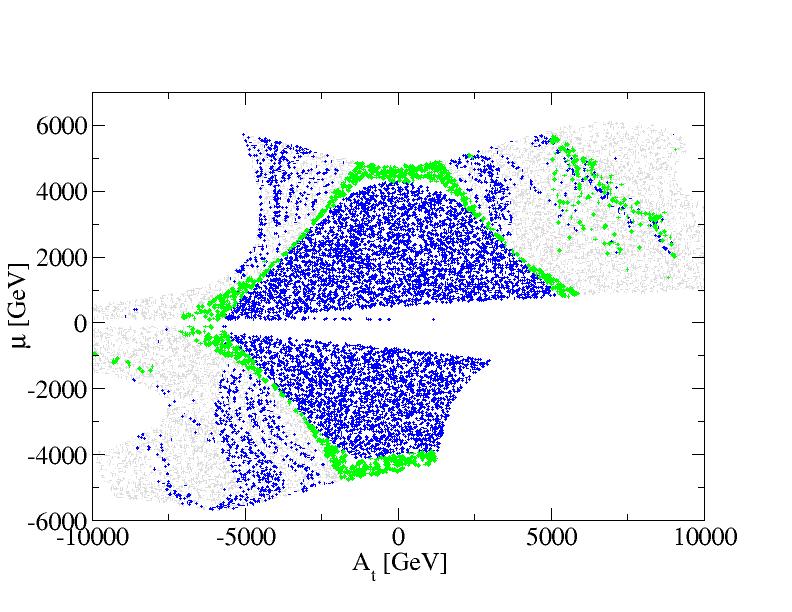} 
\caption{\it Plot of $\mu$ vs $A_t$ for  
$\tan\beta=40$ and other fixed pMSSM parameters as described in 
the beginning of Sec.\ref{largetanbeta}.  
        The color codes for different vacuum stability 
conditions are same as that of Fig:~\ref{fig:mu_scan_1}.
The central zone of
stable states and the surrounding strip of long-lived states 
include the generic region 
of pMSSM parameter space, that is characterized by relatively smaller  
value of $|\mu|$. 
There exist pockets of long-lived states quite distant from 
stable states in the zone much away from the central region.
A significant region of parameter space for large and positive values of 
$A_t$, particularly for $\mu<0$ is eliminated via 
${\rm Br}(B \rightarrow X_s \gamma)$ and ${\rm Br}(B_s \rightarrow \mu^+ \mu^-)$ limits (see text). 
}
\label{mu_at40_fig}
\end{center}
\end{figure}

\begin{figure}[ht!]
     \begin{center}
        \subfigure[]{%
            \label{fig:mh_mu40}
            \includegraphics[width=0.45\textwidth]{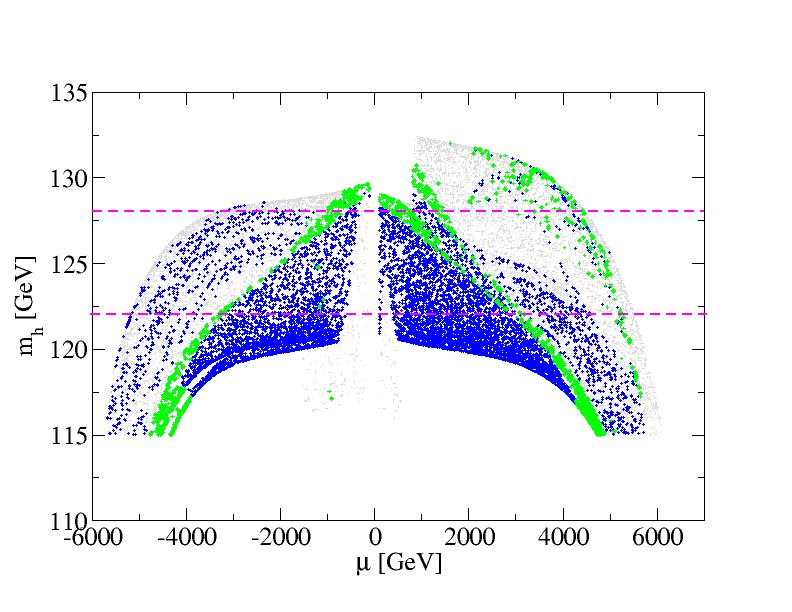}
        }%
        \subfigure[]{%
           \label{fig:mh_at40}
           \includegraphics[width=0.45\textwidth]{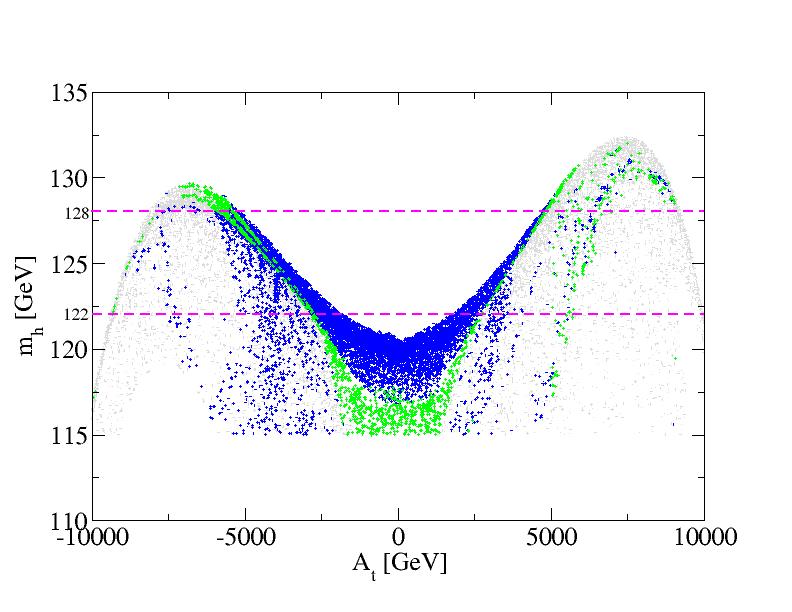}
        }
    \end{center}
    \caption{ \it
Fig.\ref{fig:mh_mu40} shows the result of scanning over $A_t$ for 
$\tan\beta=40$ in the 
plane of $m_h-\mu$ for the fixed pMSSM parameters as described in 
the beginning of Sec.\ref{largetanbeta}.     
        The color codes for different vacuum stability 
conditions are same as that of Fig:~\ref{fig:mu_scan_1}.
As in Fig.\ref{fig:mh_mu} we also identify two distinct regions 
namely Region -I and Region-II of long-lived vacua corresponding to small 
and large $\mu$ respectively for maximized $m_h$ (see text). 
Fig.\ref{fig:mh_at40} shows the result in the 
plane of $m_h-A_t$.  Here $A_t$ spans a larger zone compared to the case of 
$\tan\beta=20$ of Fig.\ref{fig:mh_at} (see text). 
     }%
   \label{fig:mh_mu_at40}
\end{figure}

The role of $\tan\beta$ in studies related to vacuum stability is important 
via its effect on the scalar potential as well as due to its influence on the 
radiative corrections to the mass of the Higgs boson, specially for large 
$\mu$ scenarios. 
In the context of Eq.\ref{sbottomeqn} the 
sbottom and even the stau loop contributions become important for 
large values of $\mu \tan\beta$ and it is 
revealed that these have negative contributions to $m_h$ that can potentially 
reduce $m_h$ below the lower limit of Eq.\ref{higgsrange}.  
Thus in this part of our work with $\tan\beta=40$ we choose a larger 
value (3 TeV) for 
the third generation of squark mass parameter in order to respect 
the Higgs mass limits, while keeping the same values of other 
pMSSM parameters of Sec.\ref{largemuscansection}. 
The combined sbottom and stau loop 
contributions typically amounts to 10-15 percent within the range of 
Higgs boson mass of Eq.\ref{higgsrange}. 
Along with the Higgs fields, we again allow $\stopl$,$\stopr$,$\sbottoml$~and
$\sbottomr$~ to acquire non-zero \Vevs~ and 
choose the following ranges for $\mu$, $A_t$ and $A_b$.
\begin{eqnarray}
\label{mu_at_scan40}
%\centering
-10~\tev~\leqslant~A_t~&\leqslant&~10~\tev, \ \notag\\
-6~\tev~\leqslant~A_b~&\leqslant&~6~\tev,\ \notag\\
-7~\tev~\leqslant~\mu~&\leqslant &~7~\tev\  . 
\end{eqnarray} 
As before we consider vanishing trilinear couplings except $A_t$ and $A_b$. 
Compared to the case of $\tan\beta=20$, here the range of $\mu$ giving valid parameter point becomes 
smaller because of the Higgs mass limits as mentioned above.  
Similar to Sec.\ref{large_mu_scenario} we impose B-Physics constraints of 
Eqs.~\ref{bphysics_constraints1},~\ref{bphysics_constraints2} on the 
resulting spectrum. \\ 
Fig.\ref{mu_at40_fig} shows the result of parameter scanning in the plane of 
$\mu-A_t$ where we have not used the constraints of Higgs mass limits.  
A significant region of parameter space for large and positive $A_t$, 
particularly for $\mu<0$ is eliminated via ${\rm Br}(B \rightarrow X_s \gamma)$  limits.  
This typically happens due to cancellation between the chargino and the combined 
contributions of $t-W$ loop from SM along with charged Higgs loops.  
This may reduce the above branching ratio to values smaller than the 
lower limit of Eq.\ref{bphysics_constraints1}.
This is consistent with the expected result of ${\rm Br}(B \rightarrow X_s \gamma)$ for 
$\mu A_t<0$\cite{Bhattacherjee:2013vga,Haisch:2012re,bsgammaextra}.  
Similarly a large amount of parameter zone of the same quadrant is 
eliminated via ${\rm Br}(B_s \rightarrow \mu^+ \mu^-)$ 
which becomes sensitive for large $\tan\beta$\cite{bsmumuextra}, 
in spite of the fact that the pseudoscalar Higgs mass 
is quite large. 
We note that the above  
agrees with the analysis of Ref.\cite{Haisch:2012re} where 
${\rm Br}(B_s \rightarrow \mu^+ \mu^-)$ 
for a large $\tan\beta $ is seen to be  enhanced for $\mu A_t<0$.
Interestingly, in contrast to Fig.\ref{fig:mu_scan_1} here 
the safe vacua are almost exclusively long-lived in the large $\mu$ and
large $A_t$ region.  \\
A similar result 
when projected into $m_h -\mu$ plane is shown in Fig.\ref{fig:mh_mu40}. 
As in Fig.\ref{fig:mh_mu} we obtain two distinct regions 
namely Region -I and Region-II of long-lived vacua corresponding to small 
and large $\mu$ respectively for maximized $m_h$ cases.  
Region-I (long-lived) is identified with $\mu \simeq 1 ~\tev $ and $m_h \simeq 127 ~\gev $ 
(small $\mu$ zone) whereas Region-II (long-lived) occurs with $4~\tev < \mu < 5.5~\tev~$ and 
$122~\gev \lsim m_h \lsim 128~\gev~$(large $\mu$ zone).
Going from $\tan\beta=20$ to $\tan\beta=40$ we see that 
$|\mu|$ cannot assume very large values because this would 
lead to a rapid decrease 
of $m_h$ when $\mu$ is increased, via radiative corrections from 
the sbottom and stau loops (Eq.\ref{sbottomeqn}). 
With the same parameter scan, Fig.\ref{fig:mh_at40} is similar 
to Fig.\ref{fig:mh_at} except that it 
refers to $\tan\beta=40$ along with a heavier third generation of squarks. 
Unlike Fig.\ref{fig:mh_at} here the EWSB vacuum is mostly
long-lived for large values of $A_t$ which also spans a larger range.
\begin{figure}[ht!]
\begin{center}
\includegraphics[height=75mm]{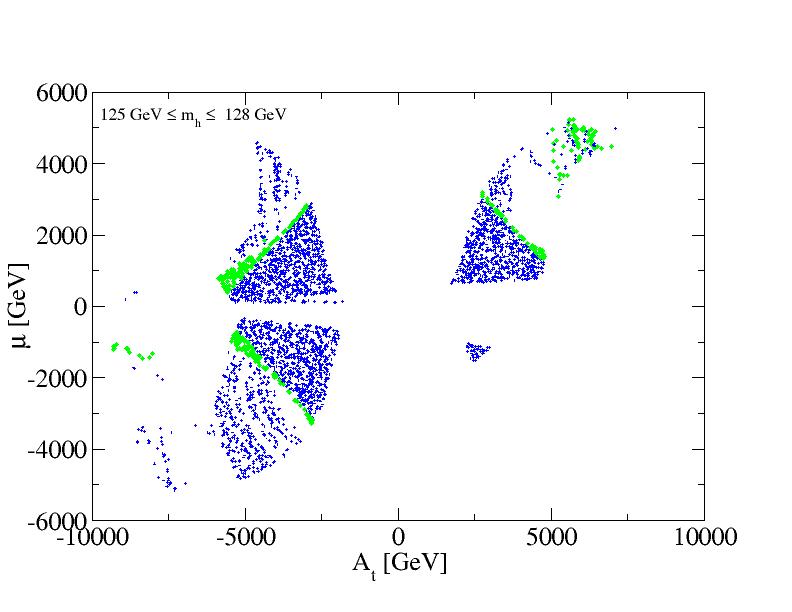} 
\caption{\it Plot of $\mu$ vs $A_t$ for 
$\tan\beta=40$ after imposing limits on $m_h$ from 
Eq.\ref{higgsrange}.   
The areas  
corresponding to stable states in different quadrants 
are appreciably shrinked for this case of a large $\tan\beta$.
The EWSB SML vacuum states in large $\mu$ and 
large $A_t$ region which are also distinctly identified 
are mostly long-lived unlike the case 
of $\tan\beta=20$. For very large values of $|\mu|$, long-lived states 
are associated with large values of $|A_t|$. The requirement 
to satisfy the limits of $m_h$ eliminates small to moderate $|A_t|$ 
zones and the regions with small $|\mu|$ and very large $A_t$ due to 
radiative corrections in $m_h$ (see text).  
}
\label{mh_mu_at40_fig}
\end{center}
\end{figure}

\noindent
In Fig.\ref{mh_mu_at40_fig}  we show the distribution 
of parameter points for stable and long-lived vacua in $\mu-A_t$ plane where
$m_h$ lies in the range of Eq.\ref{higgsrange}. The areas  
corresponding to stable states in different quadrants 
are appreciably shrinked for this case of a large $\tan\beta$.
On the other hand, the EWSB SML vacua in large $\mu$ and 
large $A_t$ region which are also distinctly isolated (green) in the figure
are mostly long-lived unlike the case 
of $\tan\beta=20$. For very large values of $|\mu|$, long-lived states 
are associated with large values of $|A_t|$. We note that the requirement 
to satisfy the limits of $m_h$ eliminates the small to moderate $|A_t|$ 
zones and the regions with small $|\mu|$ and very large $A_t$ due to 
radiative corrections. The latter 
combination enhances $m_h$ to cross the upper bound of Eq.\ref{higgsrange}. 

\begin{figure}[H]
\begin{center}
\includegraphics[height=90mm]{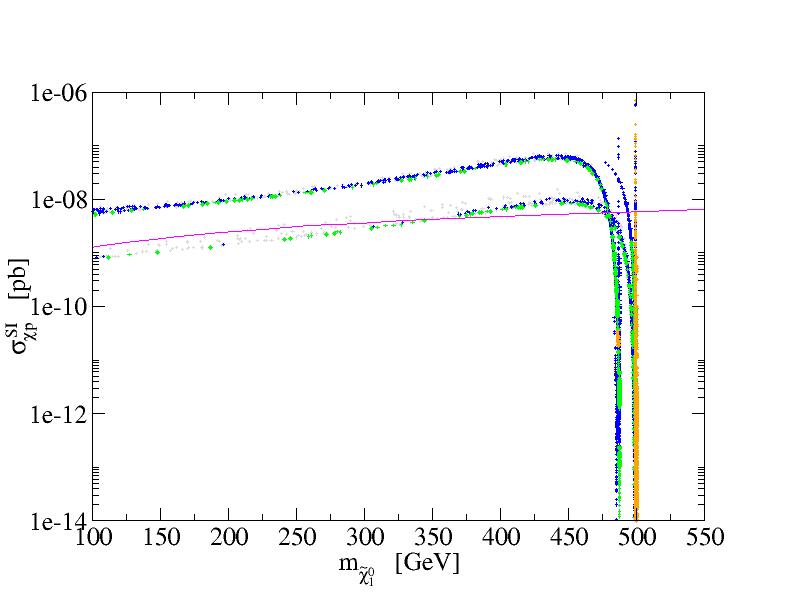} 
\caption{\it Plot of $\mlspone-\sigmasip$ for $\tan\beta=40$ of the analysis of 
Sec.\ref{largetanbeta}. See text and caption of Fig.\ref{fig:sigma} for further details.
}
\label{lux40_fig}
\end{center}
\end{figure}

We now briefly discuss the compatibility of our analysis of  
$\tan\beta=40$ with 
dark matter related data for the relic density limits from 
WMAP\cite{Hinshaw:2012aka}/PLANCK\cite{Ade:2013zuv} and 
spin-independent direct detection $\lspone-p$ cross-section measurement from 
LUX~\cite{Akerib:2013tjd}.  Fig.\ref{lux40_fig} shows the 
variation of $\sigmasip$ with $\mlspone$. The region above the pink line is excluded by LUX
results.  The regions with adequate $\varOmega_{\lspone}h^2$ 
as constrained by Eq.\ref{omega_eqn} is shown in orange dots. The other 
features of Fig.\ref{lux40_fig} are similar to what is described for 
Fig.\ref{fig:sigma}. 
\subsection{Muon $g-2$}
Finding both large $\mu$ and small $|\mu|$ regions corresponding to Region-II and Region-I respectively 
as referred before, to be valid long-lived vacuum states we immediately like to relate this 
to the issue of satisfying the constraint from 
Muon Anomalous Magnetic Moment. At one-loop level, the 
supersymmetric contributions to $a_\mu$
with $a_{\mu}=\frac{1}{2}(g-2)_{\mu}$\cite{Jegerlehner:2009ry,muong1}
originate from chargino-sneutrino and 
neutralino-smuon loops.  Large contributions may come from 
neutralino-smuon loops when $\mu$ is large along with 
smaller slepton masses\cite{Endo_muong}. On the other hand, for small $\mu$ 
zones smaller masses of charginos would increase  
the supersymmetric contributions $a_{\mu}^{SUSY}$ to $a_{\mu}$.  
The experimental
data ($\equiv a_{\mu}^{exp}$)\cite{Bennett:2006fi,Roberts:2010cj} 
 differs significantly from SM prediction 
($\equiv a_{\mu}^{SM}$)\cite{Hagiwara:2011af,Nyffeler:2013lia} leading to 
the following result where errors are added in quadrature. 
\begin{align}
a_{\mu}^{SUSY}=\Delta a_{\mu} = a_{\mu}^{exp} - a_{\mu}^{SM} = (29.3\pm 9.0)\times 10^{-10} . 
\end{align}
At the level of 2$\sigma$ one has,  
\begin{align}
11.3\times 10^{-10} < a_{\mu}^{SUSY} < 47.3\times 10^{-10}.
\label{sigma2} 
\end{align}
Focusing on muon $g-2$ constraint for a $2\sigma$ limit, we now present 
Table~\ref{table-benchmark3} for four benchmark 
points, two each for $\tan\beta=20$ and $40$.  
These are associated with long-lived states 
corresponding to Region-I and Region-II as mentioned before. 
The point for $\tan\beta=20$ in Region-I corresponding 
to smaller $\mu$ satisfies $a_{\mu}^{SUSY}$ constraint of 
Eq.~\ref{sigma2} principally  because of smaller lighter  
chargino mass\cite{Endo_muong}. For the Region -I point corresponding to 
$\tan\beta=40$, there is a natural enhancement of $a_{\mu}^{SUSY}$
due to larger $\tan\beta$. As a result a larger lighter chargino mass 
could be accommodated. 
For the long-lived state associated with 
Region-II, where $\mu$ is large, smaller values of slepton masses are required  
in order to satisfy Eq.~\ref{sigma2} because $\mu$ is large.  
This is connected with the larger contribution from the neutralino-smuon 
part of the diagrams for $a_{\mu}^{SUSY}$\cite{Endo_muong}. 
The contributions from the above type of diagram is more dominant for 
$\tan\beta=20$ where $\mu$ is much larger than the case of 
$\tan\beta=40$. However, for the two points belonging to Region-II 
an increase in $a_{\mu}^{SUSY}$ due to an increase in $\tan\beta$ is counterbalanced 
by a decrease in $\mu$, since $\mu$ spans a smaller zone for $\tan\beta=40$.

Finally, we would like to comment that for the benchmark point of Region-II 
corresponding to $\tan\beta=20$, the lighter top-squark mass in particular is 
adequately light that in turn arises out of a sufficiently large $\mu$. 
The largeness of $\mu$ indeed causes negative contributions to the radiative 
corrections to Higgs boson mass via sbottom and stau loops as discussed before. 
This effectively reduces $m_h$ which on the other hand 
allows to accommodate a larger $A_t$. The latter in turn 
gives rise to a lighter ${\tilde t}_1$ via left-right mixing. Similar 
effect for $\tan\beta=40$ also holds good but is limited via smaller 
value of $\mu$ that is allowed via vacuum stability requirement for long-lived states 
with proper Higgs mass.     
\begin{table}[H]
\caption{\it Benchmark points for long-lived vacuum states}
\centering
\begin{tabular}{|c|c|c|c|c|}
\hline\hline 
Parameters  & $\in$ Region I & $\in$ Region II  & $\in$ Region I & $\in$ Region II\\ [0.5ex]
\hline
%a & b & c \\
$m_{1,2,3}$  & {160, 179, 1400} &   {500, 525, 1400} &  {490, 550, 1400} & {500, 525, 1400}\\
$m_{\tilde{Q}_{3}}/ m_{\tilde{U}_{3}}/ m_{\tilde{D}_{3}}$  & {\bf 2000} & {\bf 2000} & {\bf 3000} & {\bf 3000}\\
$m_{\tilde{Q}_{2}}/ m_{\tilde{U}_{2}}/ m_{\tilde{D}_{2}}$  & 1000 & 1000 & 1000 & 1000\\
$m_{\tilde{Q}_{1}}/ m_{\tilde{U}_{1}}/ m_{\tilde{D}_{1}}$ & 1000 & 1000 & 1000 & 1000\\
$m_{\tilde{L}_{3}}/m_{\tilde{E}_{3}}$ & 1000 & 1000 & 1000 & 1000\\
$m_{\tilde{L}_{2}}/m_{\tilde{E}_{2}}$ & 430 & 600 & 510 & 572 \\
$m_{\tilde{L}_{1}}/ m_{\tilde{E}_{1}}$ & 430 & 600 & 510 & 572\\
$A_{t},A_{b},A_{\tau}$  & 3500, 0, 0 & 5188.5, -2640.2, 0 & 4691.2, 0, 0 & 6273.4, -3040.7, 0 \\
$\tan\beta$ & {\bf 20} & {\bf 20} & {\bf 40} & {\bf 40} \\
$\mu$  & 1000 &    8831.0 & 1500.0 & 4940.2\\
$m_A$  & 1000 & 1000 & 1000 & 1000 \\
\hline 
$m_{\tilde g}$ & 1486.9 & 1486.7 & 1531.6 & 1531.6\\
$m_{\tilde u_L}$  & 1083.5 & 1083.2 & 1179.8 & 1107.9\\
$m_{\tilde t_1},m_{\tilde t_2}$& 1880.0, 2113.5 & 922.7, 1683.7 & 2870.1, 3088.2 & 2771.3, 3064.7 \\
$m_{\tilde b_1},m_{\tilde b_2}$  & 2035.2, 2054.8 & 1986.6, 2101.4 & 3023.6, 3060.8 & 2995.9, 3087.9 \\
$m_{\tilde e_L}, m_{\tilde {\nu_e}}$ & 432.4, 425.3 &  601.8, 596.8 & 512.1, 506.1 & 573.4, 568.3 \\
$m_{{\tilde \tau}_1},m_{\tilde {\nu_\tau}}$  &984.0, 998.0 &  838.8, 998.0 & 946.3, 998.0 & 810.8, 998.0\\
$m_{{\tilde \chi_1}^{\pm}},m_{{\tilde \chi_2}^{\pm}}$  & 177.2, 1006.4 & 524.9, 8831.7 & 548.1, 1505.4 & 524.8, 4941.5 \\
$m_{{\tilde \chi_1}^0},m_{{\tilde \chi_2}^0}$ & 159.4, 177.3 &  500.0, 524.9 & 489.4, 548.1 & 500.0, 524.8 \\
$m_{{\tilde \chi_3}^0},m_{{\tilde \chi_4}^0}$ & 1003.1, 1005.4 & 8313.4, 8313.5 & 1502.5, 1505.1 & 4940.9, 4941.2\\
$m_{H^{\pm}}$ & 1003.5 & 1001.2 & 1003.4 & 1002.7 \\
$m_H,m_h$ & 1000.0, 126.8 & 988.8, 122.1 & 1000.0, 127.5 & 999.5, 124.9\\
\hline
${\rm Br}(B \rightarrow X_s \gamma)$ &$3.67\times10^{-4}$ & $2.85\times10^{-4}$ & $3.75\times10^{-4}$ & $3.25\times10^{-4}$ \\
${\rm Br}(B_s \rightarrow \mu^+ \mu^-)$  & $3.17\times10^{-9}$ & $3.23\times10^{-9}$ & $1.85\times10^{-9}$ & $1.95\times10^{-9}$\\
$a_{\mu}$ & $11.9\times10^{-10}$ & $12.0\times10^{-10}$ & $11.8\times10^{-10}$ & $16.5\times10^{-10}$ \\
\hline
$\varOmega_{\lspone}h^2$ & 0.128 & 0.118 & 0.113 & 0.107\\%[0.8ex] 
$\rm  {\sigmachionesip}$ in pb & $3.74\times10^{-11}$ & $1.82\times10^{-13}$ & $3.92\times10^{-11}$ & $9.07\times10^{-13}$\\
\hline
\end{tabular}
\label{table-benchmark3}
\end{table}
\section{Conclusion}
It is exciting that the Higgs boson has been discovered in LHC 
with its mass around 125~\gev~ 
which is well within the MSSM predicted upper limit of 135~\gev~ or so. 
However, such a relatively heavy Higgs boson is not so friendly 
in terms of the hierarchy problem.  It is thus important to explore the 
MSSM parameter space that may still be associated with a relatively 
lighter SUSY spectra. 
The observed value requires large radiative 
corrections to the Higgs boson mass that is driven by the third generation scalars 
with large Yukawa couplings, particularly the top squarks.  
It is possible to limit the latter to become not so heavy by considering large mixing between 
the left and the right scalar components. 
Choosing a large value of trilinear coupling $|A_t|$ may induce 
large radiative corrections but this could be limited by the appearance of the 
Charge and Color Breaking Minima. 
It may be seen that the role of the higgsino mixing 
parameter $\mu$ may be important in addition to $A_t$ 
while discussing CCB constraint.
We discussed the negative corrections from sbottom and stau 
loop contributions to the Higgs mass for large values of 
$\mu \tan\beta$. An effectively reduced value of $m_h$ 
as such allows a wider zone of $A_t$, which in turn 
may reduce the top squark masses in a significant zone of 
MSSM parameter space. 

Traditionally, appearance of global CCB minima is avoided via use of 
analytic relations like Eqs.~\ref{traditional_bound1} or \ref{traditional_bound2a}
of Sec.\ref{ccbmssm} as explored in the work of 
Casas {\it et al}\cite{Casas:1995pd}.  
Ref.\cite{Bordner:1995fh,Kusenko_paper1} and recently Ref.\cite{Cohen:2013kna} 
considered 
the existence of long-lived 
vacuum states for which the transition to the global CCB minima requires 
a time larger than the age of the Universe. The above references hence used relaxed 
constraints like Eqs.\ref{relaxed_bound1},\ref{relaxed_bound2}.  
However, by searching appropriate minima numerically and computing the 
transition time, it was shown in  
Refs.\cite{Camargo-Molina:2013sta,Chowdhury:2013dka,Blinov:2013uda,Camargo-Molina:2014pwa} 
that the analytic relations  
are neither sufficient nor necessary.
We confirm their conclusions in a broader setup of 
considering all possible signs of $\mu$ and $A_t$ along with 
a wider scanning range for the above parameters using {\tt Vevacious}.
As with the above references we 
find that the long-lived states exist even in the region where 
the traditional CCB constraints are
satisfied. Moreover, long-lived states extend to the zone 
where the traditional analytical constraints on CCB are violated.  
This only shows the necessity of analyzing CCB effects 
via numerical means. 
Since the third generation of sfermions with larger Yukawa couplings 
have immediate effect on the tunneling time we consider  
non-vanishing \Vevs~ for both stop and sbottom fields.   

We find that in the generic 
region of pMSSM where $|A_t|$ and $|\mu|$ are comparatively smaller, distinct 
regions of stable and long-lived states exist irrespective 
of the signs of the above two parameters.  
Beyond the above generic pMSSM regions of stable and long-lived states 
there exist zones of dangerous vacua. 
Interestingly, we find safe vacua in a broad region of pMSSM 
parameter space with large $|\mu|$ and $|A_t|$. Furthermore, 
among the above safe vacua zones one finds long-lived states that 
fall in the interesting zone where Higgs mass radiative corrections 
maximize.  
We note that the safe vacua for a large value of $\tan\beta$ 
in large $\mu$ and large $A_t$ regions are found to be predominantly 
long-lived in nature. 
Additionally, we impose the constraints from 
${\rm Br}(B \rightarrow X_s \gamma)$ and 
${\rm Br}(B_s \rightarrow \mu^+ \mu^-)$. A large region 
of parameter space with $\mu A_t<0$ is disfavoured by the above 
constraints specially for large $\tan\beta$. The interesting 
zone of long-lived vacuum states that is associated with 
large radiative corrections to the Higgs boson mass satisfy all the 
above constraints.  
On the other hand, it is also possible to satisfy the 
constraints from dark matter experiments by appropriately choosing pockets of 
parameter space in the gaugino sector which is essentially disjoint of 
our study related to CCB vacua. For economy of computation time 
we choose a combination of closely spaced  
U(1) and SU(2) gaugino masses that satisfy the WMAP/PLANCK data via 
bino-wino coannihilations. The corresponding direct detection cross section 
$\sigmasip$ also satisfies the LUX data.  
Finally, for our analysis with large $A_t$ we find
two distinct zones of long-lived states for 
small and large values of $\mu$.  We extend the analysis by considering  
the possibility to satisfy the limits from muon $g-2$ for the above 
scenarios. It is found that limits from muon $g-2$ are satisfied via two 
distinct classes of diagrams contributing to $a_{\mu}^{SUSY}$ for the two 
cases namely small and large $\mu$.   

\section{Acknowledgment}
A.D. would like to thank the Council of Scientific and Industrial Research,
Government of India for support. U.C. would like to thank Dr. Debtosh Chowdhury 
for a brief and useful discussion. A.D. and U.C. would like to thank Dr. Ben O'Leary for 
detailed discussion about a few important technical issues regarding Vevacious.

\end{document}